\documentclass[12pt]{aastex62}

\newcommand{\kms}{\,km\,s$^{-1}$}
\newcommand{{\HII}}{H\,{\sc ii}}
\newcommand{\hetw}{{He$\,$2-10}}

\received{2 August 2018}
\revised{2 September, 2018}
\submitjournal{ApJ}

\shorttitle{ALMA CO(3-2) in He 2-10}
\shortauthors{Beck et al.}

\begin{document}

\title{Dense Molecular Filaments Feeding a Starburst: ALMA Maps of CO(3-2) in Henize 2-10}

\correspondingauthor{Sara C. Beck}
\email{becksarac@gmail.com}
\affil{School of Physics and Astronomy,
Tel Aviv University, Ramat Aviv ISRAEL 69978}

\author{Sara C.~Beck$^*$}
\affil{School of Physics and Astronomy,
Tel Aviv University, Ramat Aviv ISRAEL 69978}

\author{Jean L.~Turner}
  \affil{Department of Physics and Astronomy,
  UCLA, Los Angeles, CA 90095-1547}

%\altaffiltext{1}{}

\author{S.~Michelle.~Consiglio}
\affil{Department of Physics and Astronomy,
  UCLA, Los Angeles, CA 90095-1547}

\begin{abstract}
We present ALMA CO(3-2) observations at 0\farcs3 resolution of \hetw, a starburst dwarf galaxy and possible high-z galaxy analogue. 
 The warm dense gas traced by CO(3--2) is found in clumpy filaments that 
 are  kinematically and spatially distinct.  The filaments have no preferred orientation or direction; this may indicate that the galaxy is not evolving into a disk galaxy. Filaments appear to be feeding the active starburst; the velocity field in 
 one filament suggests acceleration onto an embedded star cluster.  The relative strengths of CO(3-2) and radio continuum vary strongly on decaparsec scales in the starburst.  There is no CO(3--2) clump  coincident with the non-thermal radio source that has been suggested to be an AGN, nor unusual kinematics.   
 
  The kinematics of the molecular gas show significant activity apparently unrelated to the current starburst. 
 The longest filament, east of the starburst, 
 has a pronounced shear of FWHM $\sim40$~\kms\ across its $\sim$50~pc width over its entire $\approx 0.5$ kpc length. The cause of the shear is not clear.
 This filament is close in projection to a 
   `dynamically distinct' CO feature previously seen in CO(1--0). 
 The most complex region  and the most highly disturbed gas velocities are in a region 200~pc south of the starburst. 
The CO(3--2) emission there reveals
  a molecular outflow, of linewidth FWZI $\sim$ 120-140 \kms, requiring an energy $\gtrsim 10^{53} \rm~ erg/s$. 
 There is at present {\it no} candidate for the driving source of this outflow.

\end{abstract}


\section{Introduction} \label{sec:intro}

Star formation in dwarf galaxies can be far more extreme than in the Galaxy; dwarf galaxy starbursts can form dense massive star clusters with high efficiency, which in turn can affect the evolution of their host galaxies. How are these starbursts triggered? From the merger of gas-rich galaxies or by accretion of circumgalactic gas?  Our picture has until now been limited by the arcsecond resolution of CO images. With the advent of  the Atacama
Large Millimeter/Submillimeter Array (ALMA) we can now study molecular gas in local galaxies at the scale of individual molecular clouds and star clusters.    
 %Filament accretion may fuel starbursts in dwarf galaxies.    By mapping the molecular 
 %gas structure with with subarcsecond resolution, we can observe starburst processes  in nearby galaxies on the scales of individual star clusters and giant molecular clouds (GMCs). 

We have accordingly used ALMA to study molecular gas in the dwarf starburst galaxy, Henize~2-10. 
\hetw\  (D= 9 Mpc; 1\arcsec=43~pc)   
is unusual among dwarf galaxies for having copious (several times $10^8 M_\odot$) molecular gas \citep{2009aap...501..495S}, comparable to the estimated stellar mass \citep{Cr17},  and  high metallicity that ranges from $\approx0.5~Z_\odot$ to near-solar \citep{2002AJ....123..772V}.   
\hetw\ hosts an active starburst a few Myr old  that was probably triggered by the merger of two dwarf galaxies \citep{1995AJ....110..116K, 2002AJ....123..772V, 2005aap...434..849H}.
  The starburst has formed more than 100 Super Star Clusters (SSCs) \citep{2000AJ...120.1273J}.
   SSCs are observed in \hetw\ at all stages of their evolution, from the youngest which are deeply embedded in molecular clouds and can be seen only in the radio and infrared \citep{2001AJ....122.1365B, 2002AJ....123..772V, 2003ApJ...597..923J}, through the less deeply embedded that can be seen in the near-infrared, to those that have emerged from the cloud to be seen in optical  and UV emission \citep{1994ApJ...423L..97C, 2000AJ...120.1273J, 2003ApJ...586..939C}.  Observations of H$\alpha$ and other nebular lines indicate turbulent conditions in the ionized gas and kpc-scale outflows \citep{1999aap...349..801M, Cr17, 2010aap...520A..82C}.
  
The youngest and most embedded clusters in \hetw\ are less than 5 Myr old \citep{2003ApJ...586..939C} and are found in the area \citet{1994ApJ...423L..97C} labeled the ``central starburst" or region A.   In infrared and radio images this central starburst is seen to comprise two large clouds, each holding $\approx10^7 M_\odot$ molecular gas and a few $10^6 M_\odot$ of young stars  
\citep[about 10 times the molecular
mass and 100 times the young stellar population of the Galactic star formation region 
W49,][]{2015ApJ...814...16B}.
These giant molecular clouds will be ionized, enriched, heated and 
eventually dispersed by the young stellar population they are creating. 
 The molecular clouds have previously been studied with single dish
 \citep{2001AJ....121..740M} spectra in CO(3--2) and 
mapped interferometrically in CO(1--0) and CO(2--1) with OVRO \citep{1995AJ....110..116K} and the Submillimeter Array \citep{2009aap...501..495S}, and in high density tracers HCN and HCO+ with ALMA \citep{2018ApJ...853..125J} at arcsecond resolutions. However, while those studies could determine the global state of the gas they could not get down to the scale of the embedded clusters, which requires subarcsecond resolution. 

We have  
obtained ALMA observations of \hetw\ at 0\farcs3 resolution in the CO(3-2) emission line at 345 GHz.  
The J=3 level of CO lies 33K above the ground state and typically has critical density for excitation 
$n_c\approx 2-3\times10^4 ~\rm cm^{-3}$; this density regime probes  the dense, warm gas 
associated with star-forming clouds.  
These maps trace the dense molecular gas with  3\kms~velocity resolution and beam $\approx12~\rm pc$, 
far higher than previous observations.   
      
  \section{Observations}
\hetw\ was observed with ALMA in the C40-4 array on 15 November 2016 in the program 2016.1.00492 (S. Beck, PI). The flux and phase calibrators were J0538-4405 and J0846-2607 and the pointing center was 08:36:15.134, -26;24;33.768 (all coordinates are ICRS).  Data cubes in 4 spectral windows, one holding the red-shifted line and the other three continuum, were created. Each cube has beam size  0\farcs3$\times$0\farcs27 
and 109 spectral planes, each 1.5 MHz wide (1.3 \kms).  The data were calibrated and imaged in the 
pipeline and analyzed in CASA 4.7.0 and AIPS.  The primary beam at 345 GHz is ~25\arcsec, FWHM;
edge affects become obvious at diameter ~18\arcsec.  The r.m.s. noise in an off-line 
channel, off-source, was $\sim2$~mJy/beam. Continuum was subtracted before imaging.
The integrated intensity image (MOM0 map) was produced by summing all emission $>\pm 1.5\sigma$
in the channel maps. The rms in the MOM0 image is $\sim 0.15$~Jy~\kms\ toward the center of
the image. Intensity-weighted velocity  (MOM1) and dispersion  (MOM2) maps
were produced using a cutoff of $\pm 4\sigma$ in the channel maps.

%The ALMA maps may not detect all the CO(3-2) emission first, because they are insensitive at scales of  $\theta_{max}\approx2.8\arcsec$ or 120 pc, and second because  the gas may be in associations of small dense clouds that are below our detection limit.   
 We can estimate the fraction of the total warm dense gas detected by these interferometer maps by comparing the current interferometer map to \citet{2001AJ....121..740M}'s single dish measurement.  \citet{2001AJ....121..740M}
 found $16.6 \rm K \pm 0.6$ \kms~ in the central position of their 22\arcsec~ beam,  which corresponds to $1000 \pm 200$~Jy~\kms\ in a 
 22\arcsec\ primary beam similar in size to the ALMA beam.  
The CO(3-2) flux summed over a 15\arcsec\ field in the ALMA maps
 is $850\pm 100$~Jy ~\kms, where the uncertainty is dominated by the estimated 10\% uncertainty in the absolute flux
 calibration and not by the receiver noise.  Within the uncertainties the ALMA maps appear to detect most ($> 70$\%) of the CO(3--2) emission
within the ALMA primary beam in \hetw.
   
\section{Dense Molecular Gas in Clumpy Filaments Fueling a Starburst} 
The ALMA CO(3--2) integrated intensity map of \hetw\ is shown  in Figure 1 overlaid on a V band HST image. 
\begin{figure}
\begin{center}
\includegraphics[width=5in]{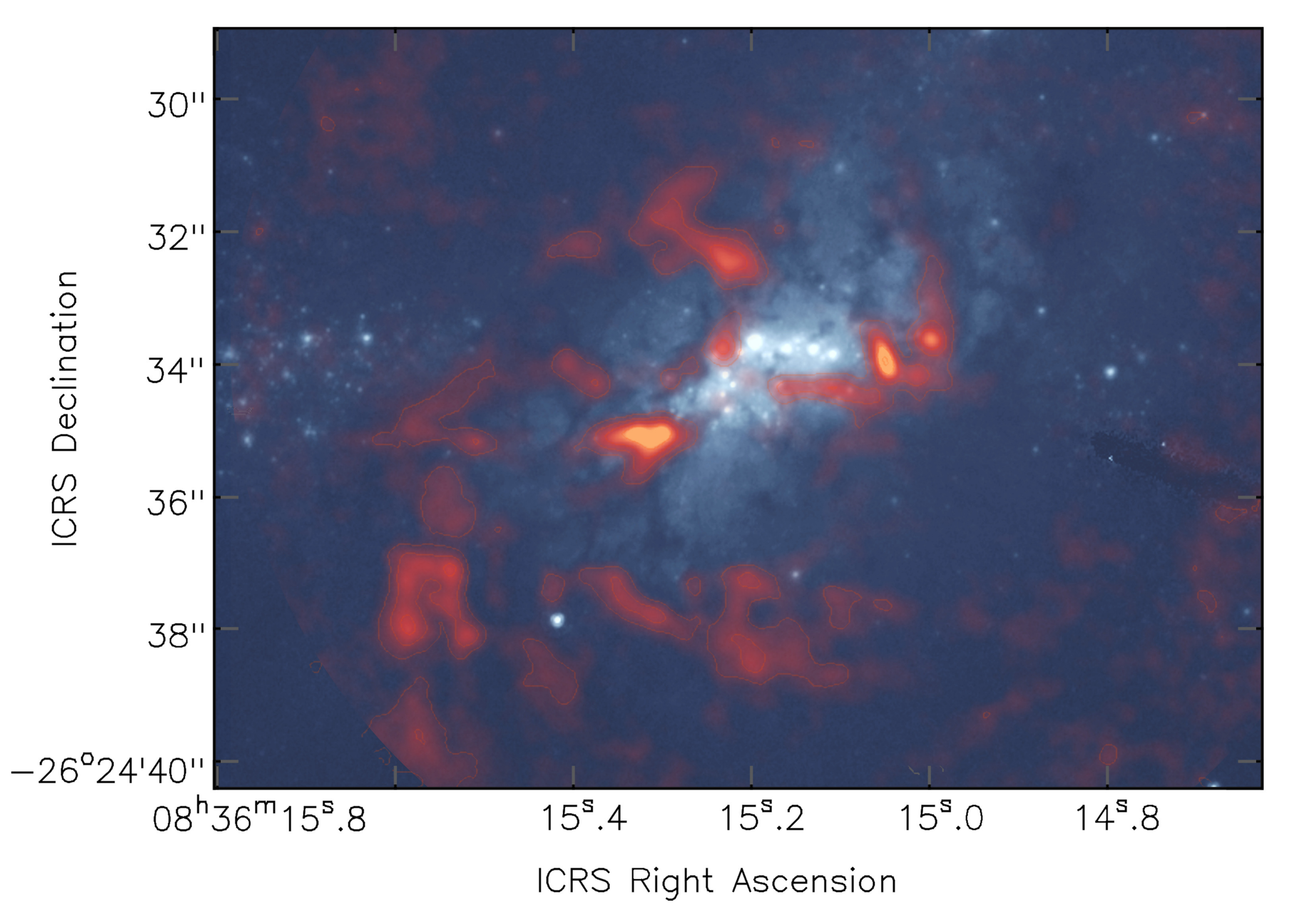}
\caption{Molecular gas and stars in He 2-10. The CO(3--2) integrated intensity (moment 0) map
is shown in red (the CO map is shown in more detail in Figure 2) overlaid on an  
HST-WFC/ACS F555W image of He 2-10 from the Hubble Legacy Archive in grey scale. The
factor of two higher noise in the outer regions of the map is due to the correction for the primary beam response. The beam of the CO observations is  0\farcs3$\times$0\farcs27.  Registration
with the HST image is limited by the uncertainty in the HST coordinates, which is 0\farcs5; however, the alignment of CO
 and dust lanes suggests that this registration of radio and optical images is good to $\lesssim 0.2$\arcsec.} 
\end{center}
\label{fig:HSToverlay}
\end{figure}
 Figure 2 displays the CO moment 0  in contours overlaid on a narrowband 658nm HST image, and Figure 3 marks the different structures with the labels which will be used for them throughout the paper.  
Comparison to the
underlying  stellar image reveals that most of the CO is found in dark dust lanes, indicating
that these clouds are on the near side of the galaxy.  
Extended filamentary structures  of CO-emitting gas are seen in the east, south, west, and center of the 25\arcsec\ 
($\sim 1$~ kpc) primary beam.  The ALMA image does not cover all of the 
CO(3-2) emission in \hetw. 
 Most of the CO is located in the East filament, a winding structure of width 0\farcs5-1\arcsec (20-40 pc) east of the bright starburst region.  We identify this filament as the inner portion of an
 extended CO streamer detected in the lower resolution CO(1-0) maps of 
\citet{1995AJ....110..116K}.   The CO(3-2) emission takes the form
of clumps instead of more extended clouds in part because the line selects dense and warm gas. 

\begin{figure}
\begin{center}
  \includegraphics[scale=0.1]{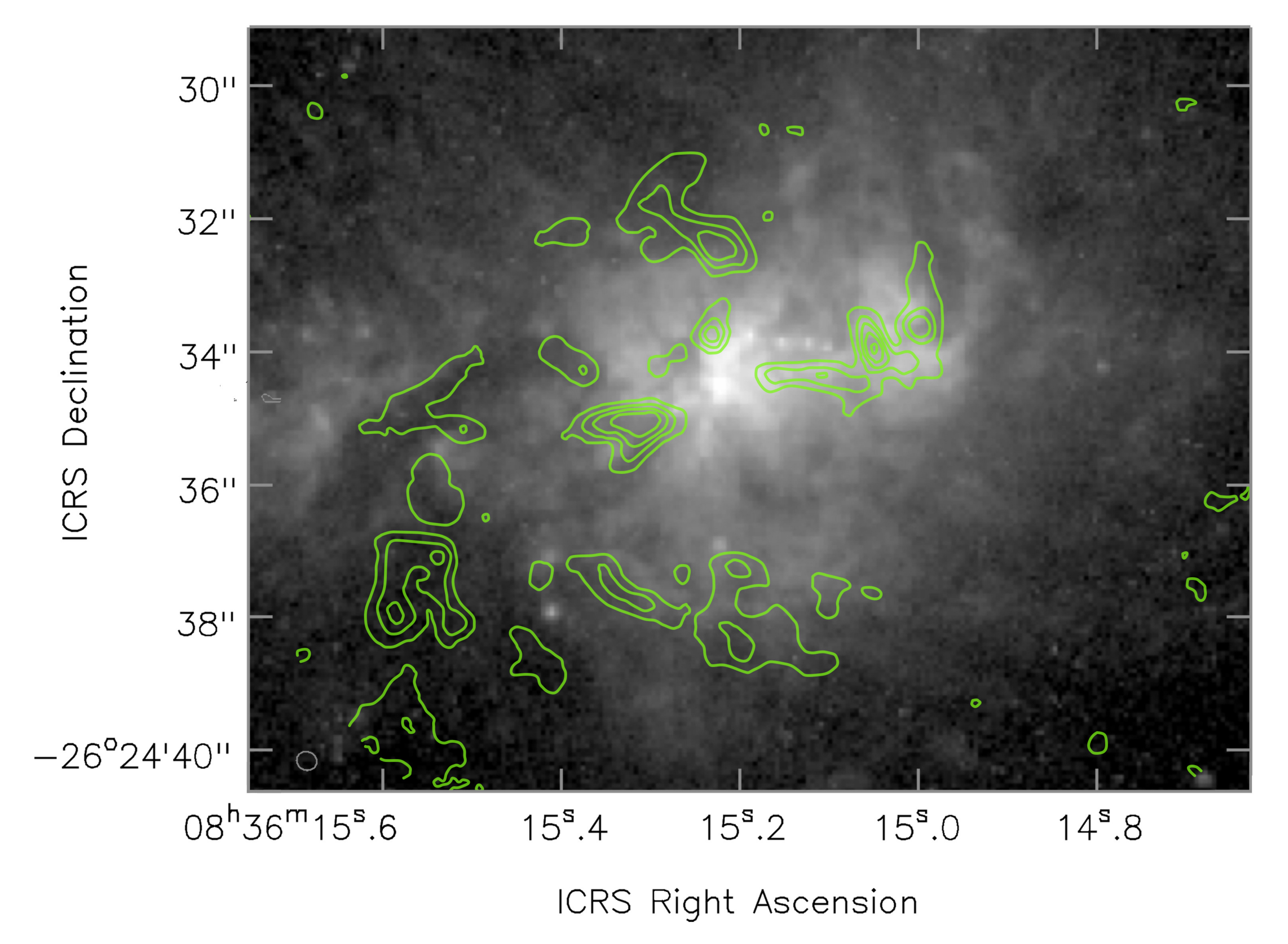} 
    \caption{The CO(3-2) moment 0 map in green, contoured at integer multiples of $8\times10^5$(Jy/bm)hz,  overlaid on an  
HST-WFC/ACS F658W image of He 2-10 from the Hubble Legacy Archive in grey scale. The CO beam is  0\farcs3$\times$0\farcs27.  }  \label{fig:labels}
 \end{center}
\end{figure}

\begin{figure}
\begin{center}
\includegraphics[scale=0.2]{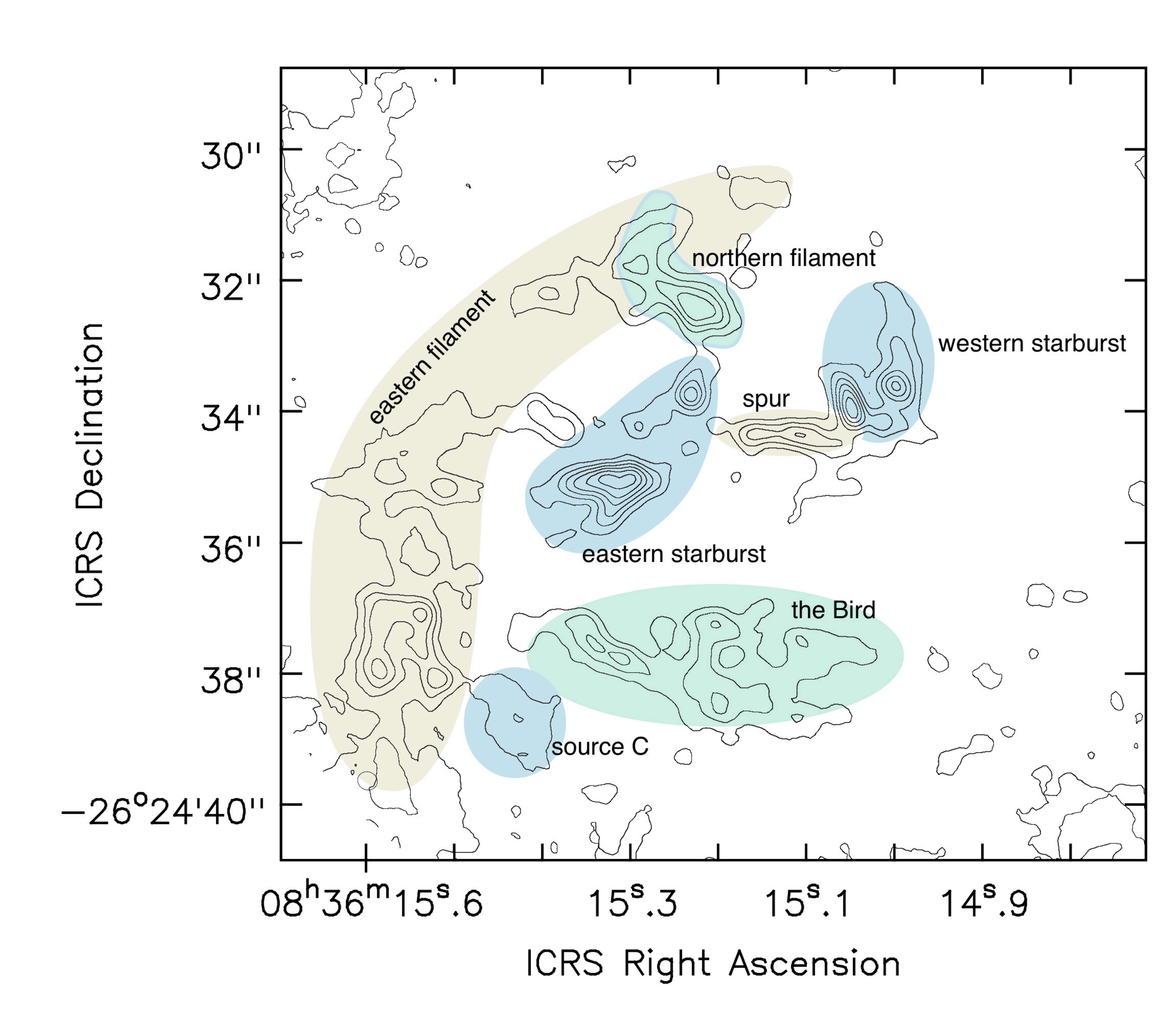}
\caption{ The CO(3-2) moment 0 map; the seven regions that will be discussed in the paper are marked and labelled.}
\end{center}
\end{figure}

 \subsection{Molecular Gas and the Starburst }

     The starburst of \hetw\ has been extensively studied in the radio and infrared 
\citep{2015ApJ...814...16B,2007AJ....133..757H, 2003ApJ...597..923J} and optical \citep{1997ApJ...488..652M}. 
The current starburst
is occurring in two giant clouds or cloud associations with total extent $\sim$4\arcsec\ , $\approx190$ pc, which are offset in velocity by 
$\sim$30\kms. These two regions hold 
$\approx6-9\times10^6 ~M_\odot$ of young OB stars \citep{2015ApJ...814...16B} concentrated in embedded clusters. In optical and near-infrared images the western starburst region is very faint and has no internal structure, while the eastern starburst region is brighter and has an arc of clusters along its north edge, as shown in Figures~1 and 2.   

The high resolution CO(3-2) images show that the active star-forming regions are themselves parts of more extended molecular filaments. We now discuss conditions of the molecular gas in the starburst and the relation of the embedded clusters to the filaments and clumps.

\subsubsection {CO(3--2) Excitation: Indications of Hot Molecular Gas in the Starburst}  
Excitation conditions can cause CO(3--2) emission to appear quite differently from the low J lines.
The 33K excitation temperature of the J=3 rotational level is slightly warmer than typical GMC temperatures, although 
the observed 
 range of temperatures in star-forming clouds extends from $\sim 20$K 
 to several hundred degrees. 
 In the average interstellar radiation field 
 CO(3--2) is expected to be strong in dense gas,  $n\gtrsim 2-3\times 10^4~\rm cm^{-3}$. 
 In the much higher radiation
 fields near O stars, where the gas can be  hot,  CO(3--2) can be subthermally
 excited and still very bright  even in lower density gas. 
So CO(3-2) can be bright in either dense/warm or 
hot/moderately dense gas, and both are conditions likely to be found in star formation regions. Without 
 some constraint on excitation, such as an additional CO line, CO(3--2) cannot  be  simply scaled to
 derive the CO(1--0) brightness, or used to find masses with the CO conversion factor $X_{CO}$. 
  
 Comparison to lower J CO lines is difficult because existing images 
have much larger beams: 
the CO(2-1) map of \citet{2009aap...501..495S} from the Submillimeter
Array had a $1\farcs9\times1\farcs3$ beam,  
more than 30 times larger than is ours; 
also \citet{2018ApJ...853..125J}'s maps of dense molecular tracers have a similar beam of $1\farcs7 \times1\farcs6$.
This discrepancy in resolution  means that we cannot constrain the excitations of the individual clumps. 
 We can however degrade the resolution of the ALMA CO(3--2) map
 to the \citet{2009aap...501..495S} beam size for a preliminary view of CO excitation and gas conditions over 
 larger scales. Degrading the resolution lessens the sensitivity to the dense, hot gas because the beam
 will be dominated by larger, cooler clouds. 
To distinguish these larger emission features
from the small ALMA  clumps we designate these larger features as
 `giant molecular associations' (GMAs) at $\sim$ 2\arcsec\ (100 pc) resolution; they may comprise individual
filaments, giant molecular clouds (GMCs), and/or collections of clumps and filaments. 

Figure 4 shows that convolved to 1\farcs9 resolution the ALMA CO(3-2) resembles the (2-1) of \citet{2009aap...501..495S} in that it appears as five GMAs,  corresponding to the South, East and North filaments and the two starburst regions. These GMAs show distinct variations in the CO(3-2)/CO(2-1) ratio over $\sim$ 100 pc scales.   

\begin{figure}
 \begin{center}
 \includegraphics[width=6in]{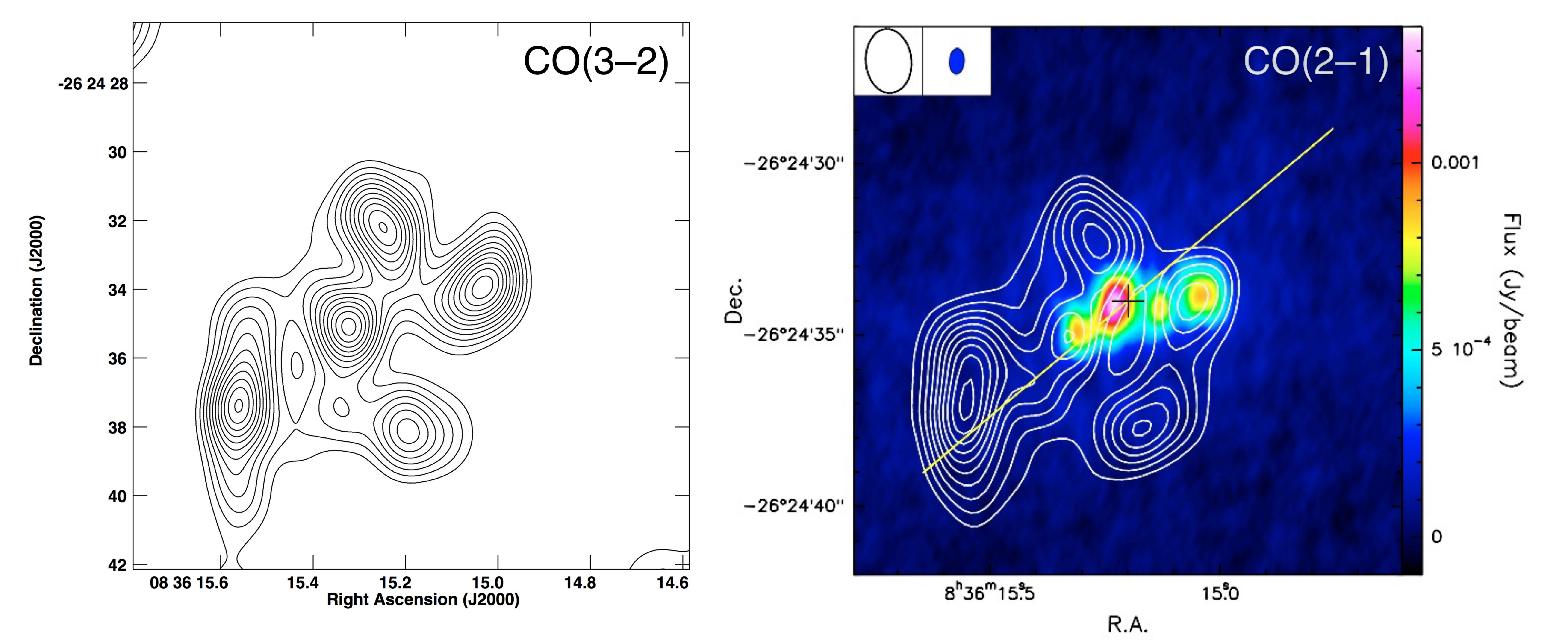}
\end{center}
    \caption{{\it (left)} The CO(3--2) ALMA map convolved to the beam of the Santangelo et al (2009)  CO(2-1) map 
    from the SMA;
    the ALMA CO(3--2) clumps and filaments appear as giant molecular associations in these much
    lower resolution maps.  {\it(right)} Figure 5 of \citet{2009aap...501..495S}. Contours are CO(2-1) emission and colors are  3.6 continuum.  Contours in both CO maps are intervals of 1.6 Jy/bm~\kms ~ starting at 4 Jy/bm~\kms,  $5\sigma$ and $2\sigma$ respectively in  the CO(2-1) map and about $10\sigma$ and $4\sigma$ for the CO(3-2).  The large-scale structure is  similar in the two CO lines, but the two GMAs closest to the starburst are significantly brighter in CO(3--2). }
 \label{fig:convolv}
\end{figure}

\begin{deluxetable}{lcccc}[t]
	\tabletypesize{\scriptsize}
	\tablecaption{CO  (3-2) and (2-1): 1.9~\arcsec Scale}
	\tablewidth{0pt}
	\tablenum{1}
\tablehead{\colhead{Cloud} &\colhead{Peak (3-2)}& 
         \colhead{Peak (2-1)} &
         \colhead{$R_{32}^{a}$}   & 
         {$T(k)^{b}$}   \\
         &
         \colhead{Jy km/s}  & \colhead{Jy km/s} \\
         &       }
\startdata
North Filament & 47  & 28 & 1.7 & 10-20\\
East Filament & 120 &  70 & 1.7 & 10-20 \\
West  Starburst  & 55 & 20 & 2.7 & $>$100 \\
East  Starburst  & 47 & 21 & 2.2 & $>$20 \\
South  Filament & 51 & 25 &  2.1 & $>$15 \\
\enddata
\tablenotetext{a}{The $\frac{(3-2)}{(2-1)}$ line ratio, with the (3-2) convolved to match  the (2-1) map. Uncertainties in
flux are 10-15\%, and thus 25\% in the ratio.}   
\tablenotetext{b} {Temperature under LTE assuming optically thick CO(3-2); if the CO(3-2) is thin these temperatures are lower limits. Temperatures are uncertain for the higher flux ratios, where they are much more sensitive to the value; CO(3--2)/CO(2-1) is not a good tracer of very warm gas. }
\end{deluxetable}

Table 1 shows the peak fluxes and $I_{CO(3-2)}/I_{CO(2-1)}$ flux and brightness ratios, $R_{32}$, in each cloud,
and the equivalent LTE temperatures for thermally excited gas.
 The estimated uncertainty in
the absolute values of these ratios due to uncertainties in absolute flux calibration of the two lines is $\sim 25\%$ and there is further uncertainty from undersampled emission.
The $R_{32}\sim 1.7$ seen for the
North and East filaments are consistent with cool 10-15 K gas. 
The  two starburst regions and the disturbed South filament  are much brighter in CO(3--2) than CO(2--1) and the formal temperatures are much higher, even on these 100 pc scales, 
although for these high ratios, the inferred 
LTE temperatures are extremely sensitive to the uncertainties for $R_{32}$.
Even given the uncertainties, the higher flux ratios imply that the 
 gas close to the star forming regions is hotter 
  than that farther away. 
  If we assume that the emission
arises in PDRs  \citep{1999RvMP...71..173H} and that  the gas is optically
thick, applying the PDR Toolbox \citep{2011ascl.soft02022P} tells us that
the line ratios of the east and north regions are consistent with gas of density $\sim 2-3\times 10^4~\rm cm^{-3}$ 
 in a standard interstellar radiation field ($G_0\sim 1$). 
The higher line ratios of $R_{32}\sim 2.1-2.7$ in the GMAs closer to the starburst can be produced by intense
 radiation fields in moderate density gas: if $G_0\gtrsim 10^5$, the density need be only $ n>1000~\rm cm^{-3}$,
 for sub-thermal emission from the hotter gas near the ionized regions.  
 Higher resolution CO maps in other rotational transitions are needed to isolate the high excitation gas in He 2-10,  
 as has been done for the low-metallicity starbursts NGC 5253 
 \citep{2017ApJ...850...54C} and II Zw 40 \citep{2016ApJ...833L...6C}. In
 these galaxies 
 ALMA measurements have made it possible to attribute  the high $R_{32}$ %CO(3--2)/CO(2--1) 
 ratios obtained in single dish observations \citep{2001AJ....121..740M,2005aap...438..855I} to the dense, warm
gas around individual star-forming star clusters.

\subsubsection{ Molecular Gas in the Starburst and the Embedded Star Clusters}
Figure 5 shows the starburst of \hetw\ in detail, including the integrated intensity image of %zeroth moment  
CO(3-2) emission (MOM0) %distribution 
and a 3.6 cm thermal radio continuum image from the VLA archive (project AJ314). The 3.6 cm radio continuum is an extinction-free tracer of \HII~regions excited by young OB stars.  
 CO(3-2) and thermal radio emission are both star formation tracers, 
 but the figure shows that  at this resolution, their distributions differ % disagreee 
to a remarkable extent.  % Going 
From west to east: %, we see that:
\begin{enumerate}
\item In the western starburst  from $\sim$ R.A. 08:36:14.95 to 08:36:15.08,  the CO and 3.6 cm coincide in a double source, as to be expected for young star clusters embedded in molecular clouds.  
The optical emission (seen in Figure 2) is very weak because of high extinction.   
\item  The `spur' of molecular gas between the two starburst regions does not show evidence of embedded stars in either the optical or the radio. 
A strong nonthermal radio point source, either an AGN \citep{2011Natur.470...66R} or SNR \citep {cr17}, is projected on the spur.
\item There are several bright star clusters in an arc along the north side
of the `spur' between R.A. 08:36:15.1 and 08:36:15.2 
that appear in the optical and $H\alpha$ images in Figures 1 and 2.  This region has very weak radio continuum, 
except possibly for the easternmost cluster, and no CO(3-2) emission. 
These appear
 to be the most evolved clusters; \citet{2003ApJ...586..939C} find that these clusters have fairly low values
 of visual extinction and are $\sim$4-5 Myr in age based
 on their UV spectra.  They are visible because they have mostly
  emerged from their molecular cocoons by photodissociating
 or blowing out  
 most of their molecular gas.   
\item  Further along the arc to the east at R.A. 08:36:15.2 to almost 08:36:15.25 
are optically detected clusters associated with three clumps of radio continuum emission.  The radio and CO emission overlap on the west of this radio triplet  but at R.A. 08:36:15.24 we see a clump of ionized gas with very weak associated CO emission.   The CO in this region may have been destroyed by high-energy radiation from the young stars or the 
non-thermal source \citep{2015MNRAS.450.4424B}. 
\item A strong radio source at R.A. 08:36:15.3 on  the eastern edge of the starburst region partly overlaps a bright clump of CO emitting gas,
and is located in a dust lane. This is apparently a young and highly obscured star cluster that has formed off-center in the molecular clump.  
\end{enumerate}

\begin{figure}
\begin{center}
  \includegraphics*[scale=0.3]{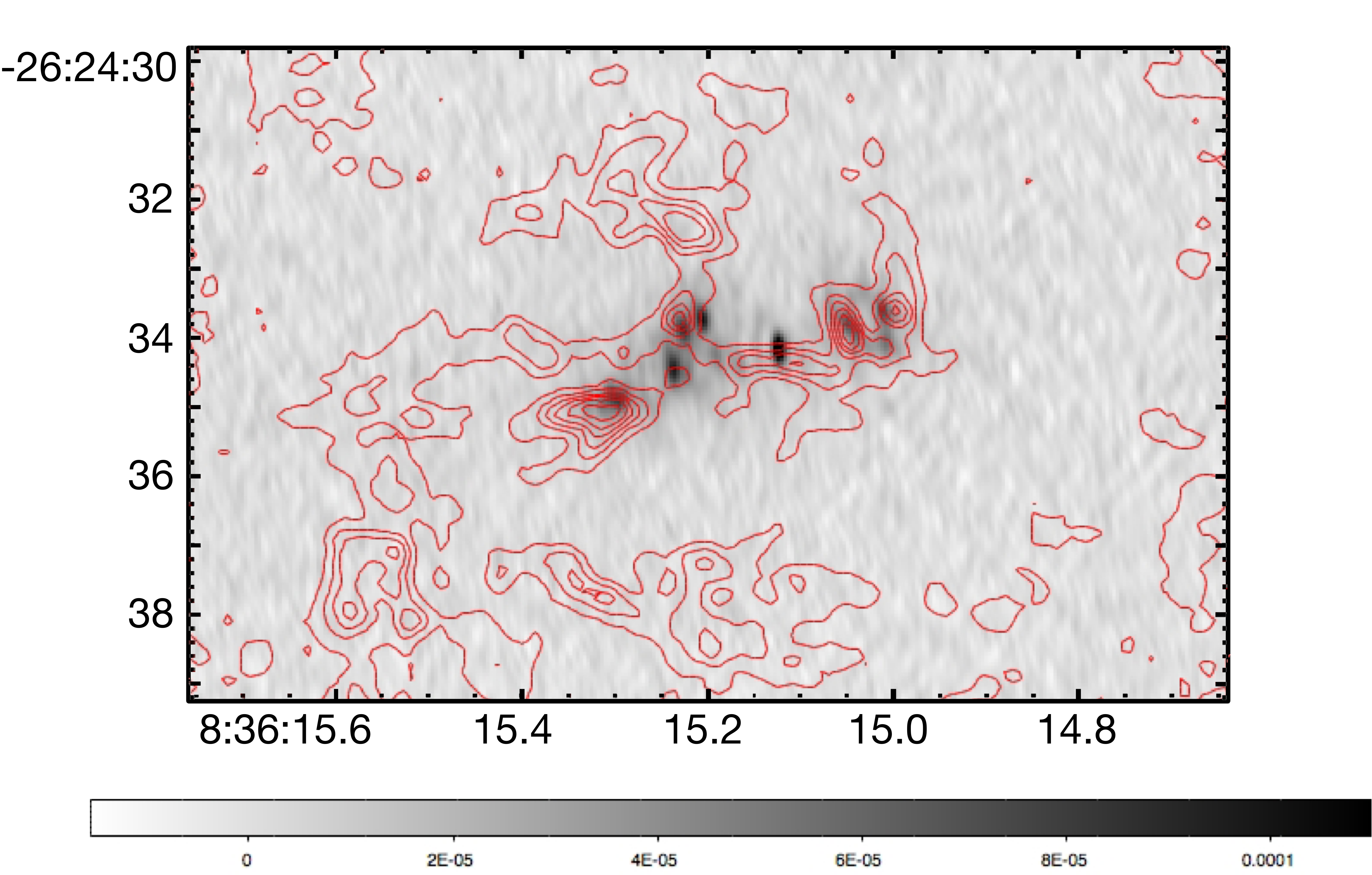}
  \end{center}
   \caption{ The starburst of He 2-10 in detail; the  CO(3-2) moment 0 map in red contours  overlaid on a
   VLA 3.6 cm radio continuum in color.  The bright point source in the middle is the non-thermal source at 08:36:15.12 \citep{2011Natur.470...66R}. The 3.6cm beam is $0.34\times0.14\arcsec$, similar to the ALMA CO(3--2)  0\farcs3$\times$0\farcs27 beam.
   Contour levels are integer multiples of  0.44 Jy/bm(\kms). }
 \label{fig:xband}
 \end{figure}

 We thus see that the distributions of molecules, radio continuum, and optical sources disagree significantly on the $\sim$ 20 pc 
scale over the eastern starburst.  The radio clumps are offset from the strongest optical and near-infrared emission, and the CO emission is further offset from the 
radio clumps.  The comparison of optical, radio continuum, and CO(3--2) suggests an inside-out model for the eastern starburst,
wherein the inner part (e.g. closest to the non-thermal source) is the most evolved and most of its gas has been cleared. The radio continuum clumps, which are less evolved star clusters because they still have gas, are on the outer edge of the optical emission region. The hot CO clumps are found even further out. 
This is in contrast to the western starburst, where the spatial distributions of the CO and radio continuum 
agree and the optical emission is suppressed by obscuration.  The differences between the two clouds are simply explained if the eastern starburst is more evolved  while the younger western has not yet  destroyed or dispersed significant amounts of gas.  This is consistent with the results of \citet{2015ApJ...814...16B} and \citet{2007AJ....133..757H} who saw ionized gas escaping the eastern starburst, 
 but not the west, through low-density channels.

What can explain the strong CO(3-2) on the spur and the variations in CO to radio continuum between clumps in the eastern starburst region?   
As noted above, CO(3--2) is not a reliable
  tracer of molecular gas mass because its brightness depends so much on temperature: in hot gas, CO(3-2) can be much brighter than expected for a given column density 
  \citep[e.g.,][]{2017ApJ...846...73T, 2016ApJ...833L...6C} .
  This may explain the strong CO(3-2) emission on the 'spur', where the gas is heated by the high ambient radiation field from the young clusters and the non-thermal source. 

\subsubsection{Masses of Molecular Clumps and Star Formation Efficiency}   
Not knowing  the true CO(3--2) excitation means we must proceed with caution to find molecular masses based on the 
 emission. For the starburst region, these will likely be overestimates, perhaps significant,  of the gas, such that the masses given by \citet{2009aap...501..495S} and \citet{1995AJ....110..116K}  are probably more reliable. However, as can
 be seen from Figures 1, 2,  and 3, much of the CO(3--2) is located far from the active star formation, and for these
 regions, such as the East filament, an
 extrapolation to CO(1--0) is reasonable.  
 The single dish CO(1-0) flux is $165\pm8$~Jy\kms\ \citep{1995AJ....110..116K}, from which we obtain a CO(3-2)/CO(1-0) flux ratio of 
 $6.1\pm1$, similar to the value of $\sim6-7$ seen on similar scales in II Zw 40  \citep{2016ApJ...833L...6C}. 
 This is equivalent to an overall brightness ratio
 of 0.7, characteristic of optically thick thermally excited gas  at 15 K.  However individual clouds may differ in excitation; 
 in particular, CO(3-2) may be brighter in warm clouds near the starburst, as suggested by the comparison to CO(2-1) discussed previously (Table 1). 
With these caveats we adopt  the CO(3-2)/CO(1-0) flux ratio of 6 and conversion factor of  $X_{co} = 4\times 10^{20}~
 \rm cm^{-2}~(K\, km\,s^{-1})^{-1}$.   We find clump 
 masses of several $10^5 M_\odot$ over areas $\approx500 \rm ~pc^2$ in the East and South filaments.  
The surface density in these clumps is over the threshold where active star formation is expected to occur  \citep{2010ApJ...724..687L}. 
 One possible explanation for
 the lack of star formation in the East filament in particular is the presence of strong shear within the filament ($\S4.1$) ,  which may, as
 in barred galaxies
 \citep{2002MNRAS.329..502M, 2013ApJ...776...70T}, prevent cloud collapse. 
 The kinematics of the East filament and the rest of the galaxy are discussed in the next section.

 A complete census of molecular clumps in He 2-10, selected by contrast over the background, is in Appendix A. 

\section{Chaotic Filament Kinematics in He 2-10 }

In stellar structure \hetw\ appears to be an early-type system \citep{2014ApJ...794...34N}.  The kinematics of \hetw\ are complex,   and provide clues to a history of interaction that has
 led to the present starburst.  The stellar component appears to be in random motion, not rotation, according to \citet{2014ApJ...794...34N} and \citet{Cr17}, who conclude that this may `preclude the existence of a dominant star-forming disk'.  The systemic velocity defined by the stars is $872$~\kms\ \citep{Cr17}, agreeing with the HI velocity of 873~\kms\ \citep{1995AJ....110..116K}.   The molecular gas is seen at moderate resolution \citep{2009aap...501..495S}  to have two velocity peaks at $\approx850$~\kms\  and $\approx890$~\kms.   \citet{2015ApJ...814...16B}  find the ionized gas associated with the starburst is also concentrated in two spatially distinct clouds, identified with the two starburst regions,  at $\approx850$~\kms and $\approx890$~\kms .  The presence of these two velocity features can create an apparent  overall velocity gradient in low and moderate resolution maps, as was reported by \citet{1995AJ....110..116K} in the first aperture synthesis observations of CO in \hetw\ .    However it should not be concluded from this apparent gradient that the molecular gas is in simple rotation;   the ALMA CO(3--2) emission shows a complex and chaotic velocity field. 

Figure 6a shows the intensity-weighted velocity (i.e., the first moment of the data cube) of the CO(3-2) emission.  CO(3-2) is seen over the same $\approx850-890$\kms~ range as in earlier observations.  But the kinematics here
are far more complex than anything seen previously.    
Instead of a gradient  in a smooth and extended gas distribution, we see distinct clumpy filaments of gas at different velocities.   (These kinematic structures may be seen more clearly in the channel maps included in Appendix B.) 
 The velocity shift from red in the northwest to blue in the southeast which has previously been interpreted as a smooth velocity gradient, is seen here to be strongly determined by a single red feature which is falling onto the western starburst (see section 4.3), and a large concentration of gas south of the starburst region appears to have an independent velocity structure.  The gas velocity is seen to change abruptly over very short distances, and different velocity features fall in the same line of sight.   The ALMA CO(3--2) maps indicate that, like the stars, the gas motions within \hetw\ are chaotic rather than
ordered.  We will now discuss the kinematics of each filament in turn. 
\subsection{The East Filament System}
The longest and best-defined filament in \hetw\ is mostly east 
of the starburst and curves west at its north end to intersect another gas component, the North filament, near 08:36:15.35, -26:24:32.   This feature is seen most clearly in the channel maps in Figure 10 (Appendix B).  This filament appears related to the CO(1-0) feature that \citet{1995AJ....110..116K} identified as a tidal tail.  The  filament is $\sim$500 pc in length and appears to extend beyond the ALMA primary beam\citep{2001AJ....121..740M}. In spite of the abundance of dense CO(3--2)-emitting gas, 
the East filament itself does not itself
show evidence of active massive star formation in the form of radio continuum emission.  The East filament largely coincides with
an optical dust lane (Figures 1 and 2) and so is on the near side of the galaxy.  
This  filament contains some of the most blue-shifted CO(3-2) in the galaxy, with velocities $\approx830$~\kms .  
It is highly clumped and the velocity varies slightly from clump to clump.  
\begin{figure}
  \includegraphics*[scale=0.6,angle=-90]{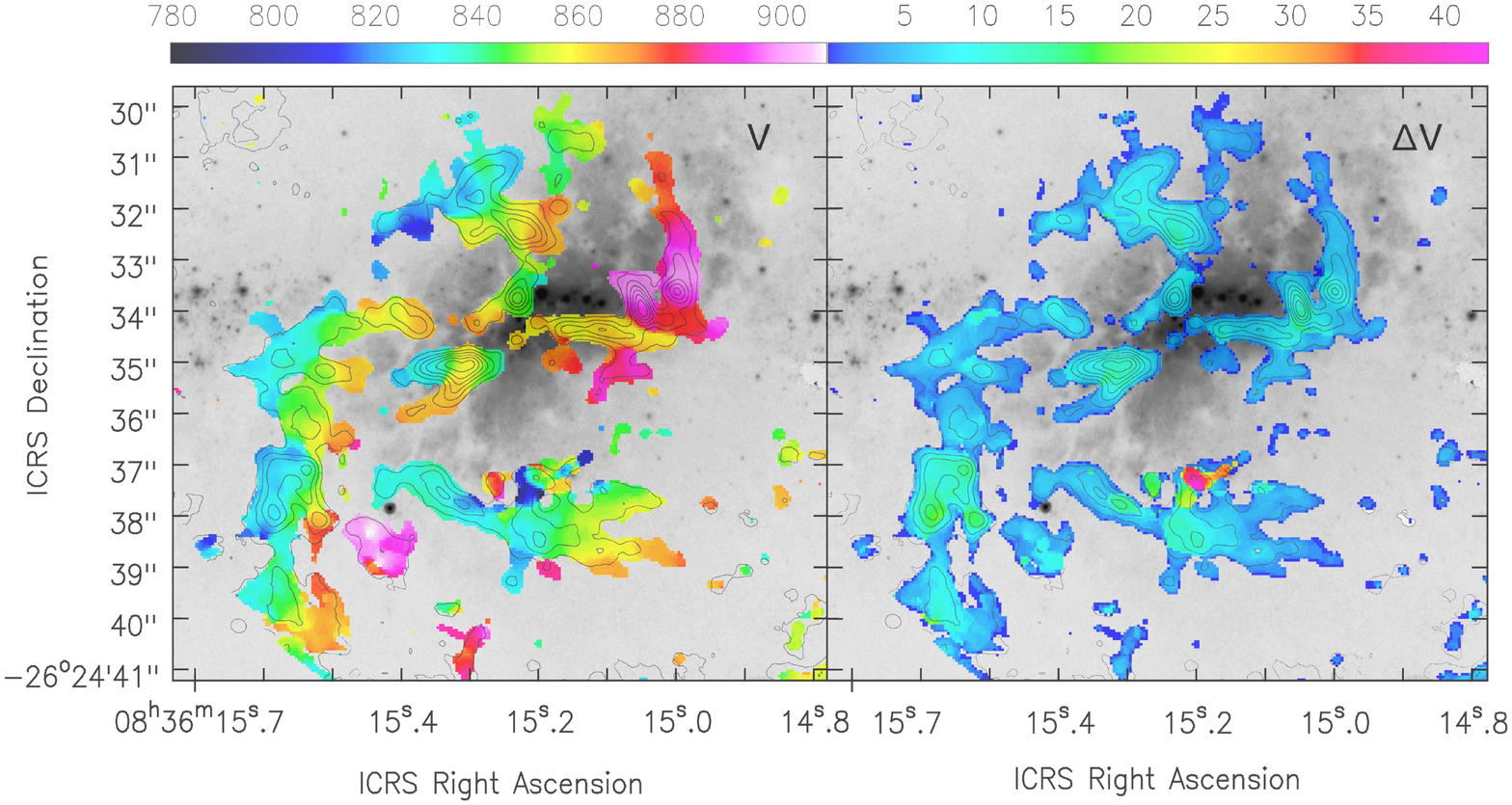}
   \caption{Kinematics of CO(3--2) in He~2-10, overlaid on the HST F555W image in grayscale.  {\it Left)} The first moment cube, giving the intensity-weighted 
   mean velocity 
   in \kms~. {\it Right)} The second moment of the cube, giving intensity-weighted velocity dispersion. The contours are from the 
    CO(3--2) integrated intensity map of Figure 1.  
   The beam is shown in lower left. Velocity scales in \kms\ are given at the top of the figures. }
  \label{fig:mom1}
 \end{figure}
 
One of the defining features of the East filament is a pronounced velocity {\it shear}. This shear is 
evident along the entire 500 pc length of the filament south of -26:24:33. The shear direction is from blue on the east to red on the west and its magnitude is 20-40 \kms~ across the filament's  30 to 50 pc width, with a maximum of $\approx55$~\kms\  near  -26:24:38.   In contrast, there is no signficant velocity gradient along the length of the filament. 
This calls into question \citet{1995AJ....110..116K}'s original identification of this structure as a tidal tail. 
In magnitude and scale 
the shear is reminiscent of that seen along spiral arms in spiral galaxies.  There is no obvious mechanism that is responsible for the shear.   The 
gas clump at -26:24:38.5  velocity $\approx 900$\kms , which we identify with the  `dynamically distinct source C' discovered by
\citet{1995AJ....110..116K} and which from its position on the dust lane is falling into the galaxy, has virial mass derived with an assumed $\rho\propto 1/r$ density distibution of $\approx4\times10^6 M_\odot$, 
not enough to cause such large scale shear by its gravitational attraction. (Further parameters of Clump C are in Table 2, Appendix A). 
  Another possible mechanism that could produce 
the shear is an hypothetical
 elongated mass along the 0.5 kpc length of the arm, a situation similar to that producing
 the shear in galactic bars and spiral arms. While this would provide a natural explanation for
 the length of the sheared filament, the hypothesized mass appears to be dark; it could be cool gas.  A final possible mechanism for the shear is that the
filament is in fact an expanding shell decelerating into a confining medium,  but there is no evidence for any explosion or other source of such expansion there.
 The shear is a consistent and
 unexplained kinematic feature
 of the East filament, and it is far from the only mysterious feature in the kinematics of \hetw.

\begin{figure}
\begin{center}
\includegraphics[scale=0.3]{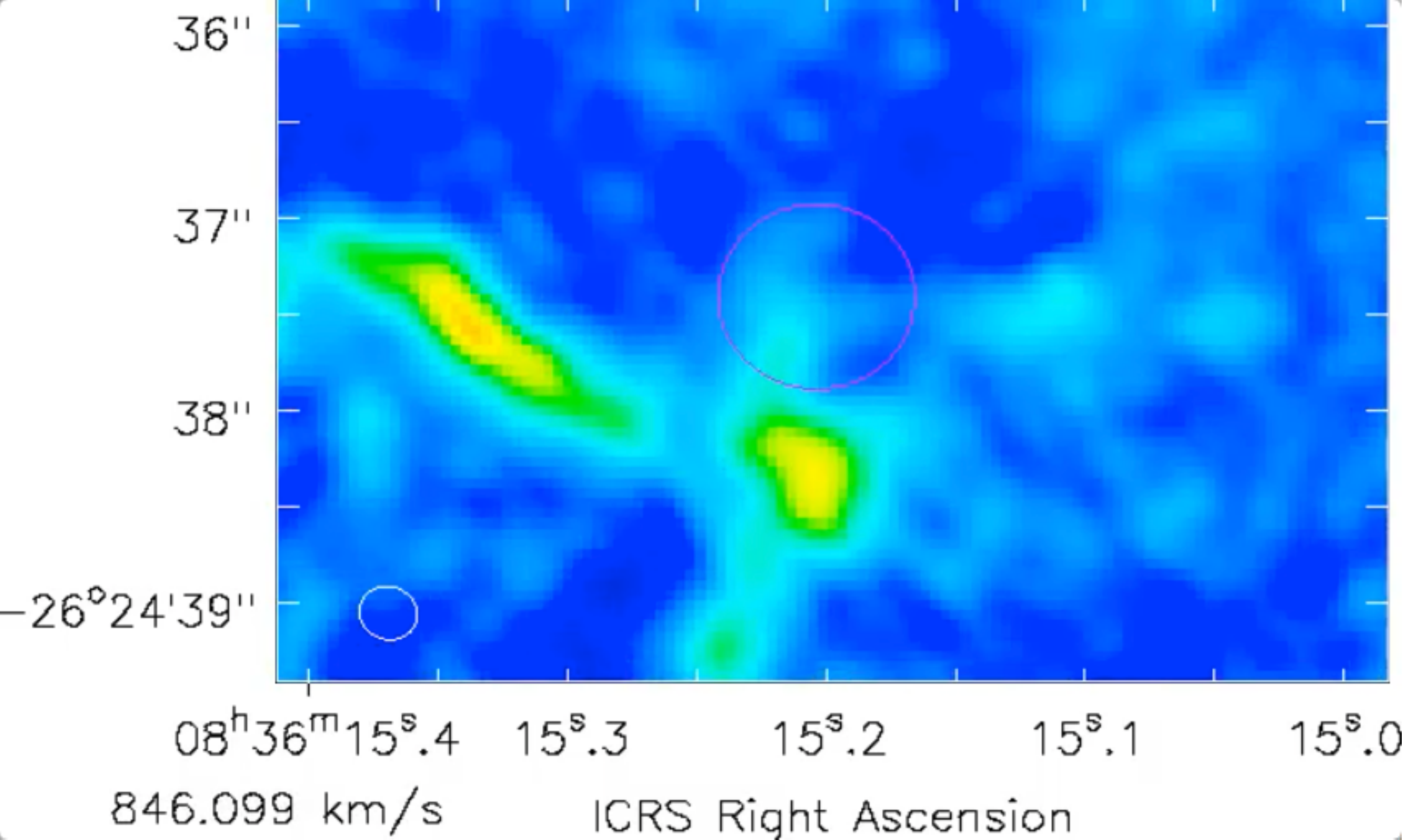}
\caption{Sample velocity channel of the ``Bird" or South Filament.  Velocity channels over the full  200 \kms  range are available online as a movie of 45 seconds duration.  The magenta circle in the animation shows the $\sim$50 pc region of very high velocity dispersion and the most extreme velocities; the more central velocities are emitted from a more extended region.}
\end{center}
\end{figure} 

\subsection{ The Southern Outflow Region: ``The Bird"} The most dynamic region in \hetw\ is the gas clump at 08:36:15.2, -26:24:37 (ICRS). 
This region stands out in the map of CO(3--2) velocity dispersion, 
(Figure 6b).  %w 
While the dispersion is $\sigma\sim$5 --12\kms\ over all the other gas in
 \hetw, in this compact region it goes up to $\sigma\sim$40~\kms. 
 In shape this region resembles a long-necked bird flying east,   with the high velocity dispersion confined to
the upper wing.   
 \citet{2009aap...501..495S} locate two density clumps in this region, 
 both very blue (820-840\kms). The CO(3-2) emission in this region is complex, consisting of multiple clumps and filaments.
The high dispersion region has emission at velocities from 875 to 920 \kms, within $<$
25-50 pc in radius. The channel maps,  animated on-line in Figure 7 and also in Figure 10, 
 show confined emission from the reddest and bluest velocities; the central velocity channels have emission extending 
radially outward.

\begin{figure}
\begin{center}
  \includegraphics[width=4in]{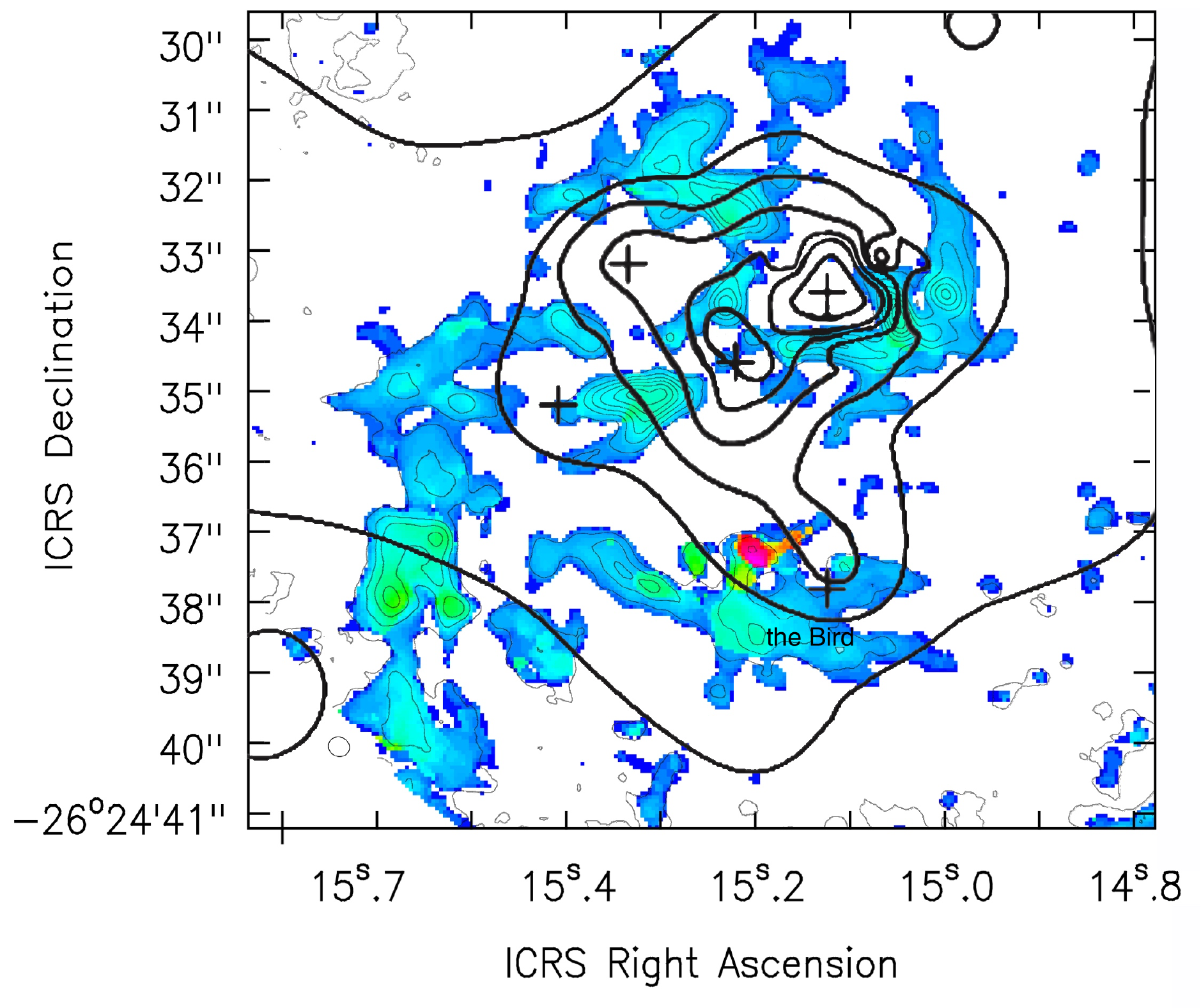}
  \end{center}
   \caption{ CO(3--2) velocity dispersion and Chandra X-ray image in He~2-10. The CO(3--2) velocity dispersion
   with the adaptively smoothed Chandra map of the 0.3-6 keV emission from \citet{2010ApJ...718..724K} overlaid.  X-ray source \#30 is on the cross near the high dispersion feature on the Bird. 
   }
  \label{fig:mom2xray}
 \end{figure}

The Bird appears to be a molecular outflow. The source of the outflow is not 
apparent. There is nothing obvious at this location 
in optical or near-infrared continuum images, nor is there radio continuum. 
There is no large-scale systematic velocity gradient in this region, so there is unlikely to be a significant concentration 
of dark mass.   
The Bird is located on the inner part of the southern arc of H$\alpha$ emission that extends to kpc scales
in an apparent  galactic bipolar outflow as detected by \citet{1999aap...349..801M}. 
However the H$\alpha$ line in the southern region, unlike in the north, is not broadened
\citep{1997ApJ...488..652M,1999aap...349..801M}, possibly because it is confined by the molecular gas. 
The Bird
is near x-ray source \#30 in \citet{2010ApJ...718..724K}, also shown in Figure 8. 
The  X-ray emission emerges from regions of low CO column and extends south  in the direction of the Bird.
Two registrations for the X-ray map have been suggested, depending on which of the two central X-ray sources lines up with the nonthermal radio
source in the center of the starburst: in the registration of \citet{2016ApJ...830L..35R} which  
places the {\it western} X-ray source on the nonthermal radio source, the X-ray spur and point 
source are closer to and almost centered on  the high dispersion region.  
 It is also notable that the molecular outflow is within 100 pc of highly redshifted Source C, although how they might
be related is unclear.
   
 We can estimate a minimum energy necessary to produce the outflow based on the
 mass and velocity. The CO flux for the Bird is $\sim 69$ Jy \kms; for CO(3--2)/CO(1--0) = 6, we
 compute a total mass for this feature based on CO(3--2) of
 $\sim 1\times 10^7~\rm M_\odot$. 
For comparison, \citet{2009aap...501..495S} find a mass of $\sim 7\times 10^7 ~\rm M_\odot$, based on lower resolution
 CO(2--1) maps (corrected to Xco = $4\times 10^{20}~\rm cm^{-2}~(K\, km\,s^{-1})^{-1}$); this suggests that the 
 molecular gas  in this region is relatively cool or relatively dilute. 
 The CO(3--2) flux
 of the smaller high dispersion region in red in the ALMA map of Figure 6 is $\sim$13 Jy \kms, thus we infer   
the mass involved in the outflow is $\gtrsim 3\times 10^6~\rm M_\odot$:
some of the more extended gas within the larger Bird region may
also be involved in the outflow but moving transverse to the line of sight, as implied by the channel maps. 
The FWZI of the CO(3--2) emission in the high dispersion region is $\Delta v \sim140$ \kms. 
If we assume that the gas is an expanding shell of material, a minimum
 energy of $\gtrsim10^{53}$ ergs is needed to produce
 the observed outflow, and more if it includes more gas moving transverse to the line of sight, as is likely. 
  This energy corresponds to 100 simultaneous SNe if all their energy
  went into gas motions. \citet{2010ApJ...718..724K}
 note that the thermal energy implied by the galaxy-wide X-ray emission requires $\sim 10^{54}$~ergs, which 
 they attribute to the effects of the starburst.
 
 While the galaxy-wide X-ray emission
 and H$\alpha$ bubbles in \hetw\ may originate in supernovae and winds within the starburst, 
the Bird region
 is $\sim 0.5$~kpc in projected distance from the starburst, and appears to be confined to a
 smaller region. 
The Bird molecular outflow may be indicative of a hidden source of energy. 
  
\subsection{The Filaments Feeding the Starburst}
 
The smallest filamentary structure is the North filament.  It might be thought part of the East filament, which it overlaps, but its line peak velocity is  the 850-860\kms\  of the east starburst cloud, more than 20\kms\ offset (this is seen most clearly in the channel maps in Appendix B).  The North filament terminates on the edge of the eastern starburst.  We argue above that the eastern starburst has lost most of its associated molecular gas; it possible that the short filament we see now is the remnant of a larger system that fed the eastern starburst and has been either used up or dispersed.  
 
\begin{figure}
\plotone{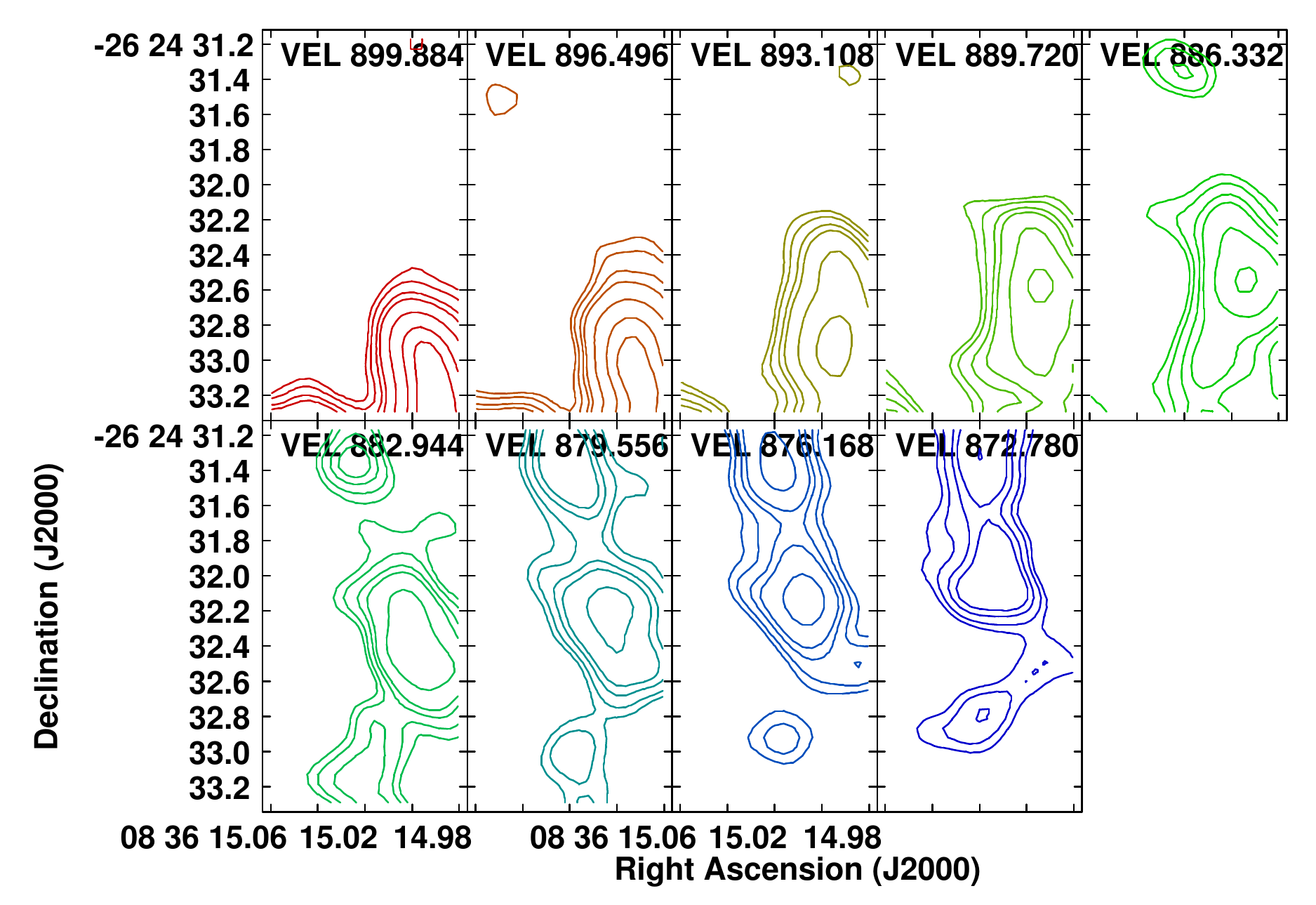}
\caption{Channel maps of the West filament showing the velocity gradient. The reddest emission is nearest the starburst,
consistent with infall onto the embedded clusters.  Contour are $2^{\frac{n}{2}}\times8.0$ mJy/beam.}
\label{westfil}
\end{figure} 

 The West filament starts north of the western starburst.  It intersects or overlaps the eastern starburst near  Dec -26:24:34.5, where both velocity components are seen in the line profile,  and re-appears south of the spur.   It has the reddest velocities in the galaxy:$\approx890-870$\kms\ . Since the West filament lies along a dust lane, it is foreground and the
 redshifted gas is falling toward the galaxy. Unlike the other filaments, the West filament
 has a significant velocity gradient along much of its length. The gradient,  shown in Figure 9, is in the direction of increasing velocity towards the starburst.   The filament appears to be accelerating into the western starburst cloud; the velocities, modulo projection effects,  are consistent with infall into a central
 mass in this immediate vicinity of a few $\times 10^7~\rm M_\odot$.  This is not far from the mass of the western starburst cloud as estimated by \citet{2009aap...501..495S} and \citet{2015ApJ...814...16B}.  This filament may still be adding fuel to the western starburst cloud.  The velocity and size of the \hetw\ West filament suggests that this is not a long-lived phenomenon; the time scale for the filament material to travel its $\approx50$~pc length into the cloud at typical velocities of $\approx30$\kms\ is only $\approx1.7\times10^6$ yr.  

We thus see in the West filament another example, with NGC 5253 \citep{2017ApJ...850...54C, 2015Natur.519..331T} and II Zw 40 \citep{2016ApJ...833L...6C} of a young starburst fed by a molecular filament. 

  \section{Discussion and Summary}
 We have presented ALMA observations of the CO(3-2) emission in He 2-10 with spatial resolution $\approx0.3\arcsec$~or 12 pc. 
 The result is a picture of the galaxy very different from that previously seen. Since the first interferometric study of this system \citep{1995AJ....110..116K} it has been thought that the starburst started from the merger of two dwarf galaxies.  Our high resolution observations add to and refine this picture.   
 The molecular gas in \hetw\ is very disordered, with clumpy filaments at many velocities and in different directions.  The longest and most massive filament is east of the starburst and is part of the structure \citet{1995AJ....110..116K} called a tidal tail. However the filament has no velocity gradient along its 0.5 kpc length;  instead it shows a strong shear across its width.  This shear may have inhibited star formation in the dense clumps of the filament. 
 
 The starburst activity is in two distinct regions that have been proposed as the cores of the merger partners.   The current observations show filaments feeding into the two starburst regions.  The  velocity gradient in the filament near the western starburst region show it is accelerating onto the young star clusters.    
 
 The gas kinematics show that the starburst is not the only mark of the disturbed past of this galaxy, or the only force shaping its development.  There are many very unusual, chaotic and dynamic kinematic features that are not all directly connected to the starburst
or any other visible feature. These include the shear in the East filament,  the gas clump in the south (source C) which is red-shifted by 70\kms~ relative to its surroundings, and also in the south the extended region which we call the Bird, where the molecular linewidths are as high as $\sim$120\kms .  The Bird appears to hold a molecular outflow of at least $10^{53}$ ergs, but with no clear cause or driving source.  

  Specifically:
\begin{itemize}
\item The CO(3-2) in \hetw\ is in 
clumps that form filamentary systems, and many of the filaments correlate with optical dust lanes. Although the implied
densities 
of the filaments are high enough for star formation, they appear at present quiescent.  
\item The total CO(3--2) flux is $\sim 850\pm 100$ Jy \kms, at least 70\% of the
single dish values. For CO(3--2)/CO(1--0) flux ratio of 6 and a Galactic conversion factor of $X_{co} = 4\times 10^{20}~
 \rm cm^{-2}~(K\, km\,s^{-1})^{-1}$,  this flux implies a total molecular gas mass, including He, of
$M_{H2gas} = 1.2 \times 10^8~\rm M_\odot$ for the central 18\arcsec\ region including the starburst; 
this may be an upper limit to the molecular gas mass if there is unusually warm CO(3-2) near the starburst. 
\item The two starburst clouds, eastern and western,
are each associated with a filament.  The two filaments appear to overlap in projection near the non-thermal radio source between the two starburst clouds. 
\item  There is a  velocity gradient along the West filament consistent with infall 
into the western starburst cloud. 
\item The non-thermal radio source  which has been variously identified as an AGN and an SNR is located in a bridge or spur of molecular gas;  it is not associated with a molecular clump.  The CO(3--2) kinematics there are quiescent, which may argue that it is an SNR rather than an AGN. 
\item The East filament shows a pronounced velocity shear of $\sim 20-40$ \kms~  across its  $\sim$50 pc width for its entire $\sim500$ pc length.  This shear may have inhibited star formation in this massive filament.   There is little or no velocity gradient along its length.  The cause of the shear is unknown. It may indicate the presence of significant
dark mass along the filament. 
\item The `dynamical source C' noted in previous CO observations has been resolved and is a gas clump of $\approx4\times10^6 M_\odot$ with velocity $\approx70$\kms~ redder than the nearby filaments.  Its relation to 
the rest of the gas, and the galaxy, is unclear.
 \item The southern region which we call the Bird has a complex velocity field consistent with a
 molecular outflow. The velocity width of the outflow is $\Delta v \sim 120-140$~\kms\ (FWZI), contained within a region
 of 25 pc extent, and it involves between $0.3-1\times 10^7~\rm M_\odot$ of molecular
 gas. The energy estimated to drive the outflow is $\gtrsim 10^{53}$~ergs and likely $10^{54}$~ergs.
There is no radio, optical, or infrared source at this location;
there is diffuse H$\alpha$ emission but the H$\alpha$ shows no kinematic irregularities here. 
The source of the molecular outflow is currently unknown. 
\end{itemize}

In short, even in this galaxy which is known as a type example of a merger-triggered starburst, there is substantial kinematic and energetic activity independent of star formation.  
The current molecular medium is dominated by dense clumpy filaments, which have complex and chaotic kinematics. There is no preferred direction or orientation of the filaments, which may show that the galaxy is not evolving into a disk system.    The complex kinematic picture of  the gas in \hetw\ cannot be simply explained by the starburst: other energy sources must be present.  The energetic events  include a large-scale shear in a massive filament, and most notably a molecular outflow of energy least $10^{53}~\rm erg$; the causes are as yet unknown but they do not appear to be related to the current starburst.   Dark forces, or matter, may be shaping this galaxy.

\acknowledgments

This paper makes use of the following ALMA data: ADS/JAO.ALMA\#2016.1.00492.S. 
ALMA is a partnership of ESO (representing its member states), NSF (USA) and NINS (Japan), together with NRC (Canada), NSC and ASIAA (Taiwan), and KASI (Republic of Korea), in cooperation with the Republic of Chile. The Joint ALMA Observatory is operated by ESO, AUI/NRAO and NAOJ. The National Radio Astronomy Observatory (NRAO) is a facility of the National Science Foundation operated under cooperative agreement by Associated Universities, Inc.  Support for this work was provided by the NSF through award GSSP SOSPA2-016 from the NRAO to SMC and grant AST 1515570 to JLT. 

\vspace{5mm} 
\facilities{ALMA}

\software{astropy \citep{2013aap...558A..33A}, 
	Starburst99 \citep{1999ApJS..123....3L,2014ApJS..212...14L},
          Cloudy \citep{2013RMxAA..49..137F}, 
          SExtractor \citep{1996aapS..117..393B}
          }

 \appendix
\section{Clump Census}
 We ran the cprops code on our CO(3-2) image cube as described in \citet{2006PASP..118..590R},
  through use of cpropstoo, which  has been optimized to run on ALMA data.
 This program works by identifying local maxima in a data cube, in this case CO(3--2), as CO clumps. 
Once a 6$\sigma$ detection is made 
in two consecutive channels, the surrounding emission
is mapped down to a two-sigma detection limit in a minimum of two consecutive channels. The minimum area required
to be uniquely associated with an individual maximum was 25 pixels, and a minimum volume of 32 pixels. Pixels are
0.053\arcsec. 
Clump detection was completed on a non-primary beam corrected image, while measurement was made on
a primary beam corrected image.  This resulted in 70 total clumps. The clumps are listed in the machine-readable table; the first 5 lines of which appear below.

The Table lists the clumps, numbered in order of right ascension, and gives the position, major axis in pc, peak line velocity, velocity FWHM, flux, luminosity and mass.. The mass is found from the CO flux using a Rayleigh-Jeans corrected brightness temperature ratio of
0.82, appropriate for 30 K gas,  for CO(3-2)/CO(1-0) $=6$ and the default cprops $X_{CO}$ of $4.35\times10^{20}
\rm cm^{-2} (K~km/s)^{-1}$. 
\nopagebreak
\begin{deluxetable}{lllllllll}[hb]
	\tabletypesize{\scriptsize}
	\tablecaption{Molecular Clumps in He 2-10}
	\tablewidth{0pt}
	\tablenum{2}
\tablehead{\colhead{ID} &\colhead{R.A.}& 
         \colhead{Dec} &
          \colhead{$L_{maj}$}  &
         \colhead{Velocity}  &
         \colhead{FWHM}   &      
         \colhead{Flux} &
         \colhead{Luminosity} &
         \colhead{Mass}    
       \\
       & & &pc& km/sec & km/sec & K km/sec  & $L_\odot$ &  $M_\odot$ }
\startdata
0	&	129.0642853	&	-26.41073036	&	25.77	&	898.1	&	45.7	&	85	&	1.6E+05	&	7.4E+05	\\
1	&	129.0639648	&	-26.41031075	&		&	892.6	&	4.8	&	1	&	2.2E+03	&	9.8E+03	\\
2	&	129.0627136	&	-26.4094162	&	14.16	&	892.3	&	30.4	&	35	&	6.7E+04	&	3.0E+05	\\
3	&	129.0635986	&	-26.41038895	&	11.10	&	892.1	&	11.7	&	7	&	1.4E+04	&	6.2E+04	\\
4	&	129.0632629	&	-26.41085815	&	12.37	&	882.0	&	15.1	&	10	&	1.9E+04	&	8.6E+04	\\
\enddata
\end{deluxetable}

 \newpage
 \section{Kinematic Channel Maps}
Velocity channels of the CO(3-2) emission are shown in Figure 10. Coordinates and intensity scale are in the last panel.   
\begin{figure}
\begin{center}
\includegraphics*[width=6in]{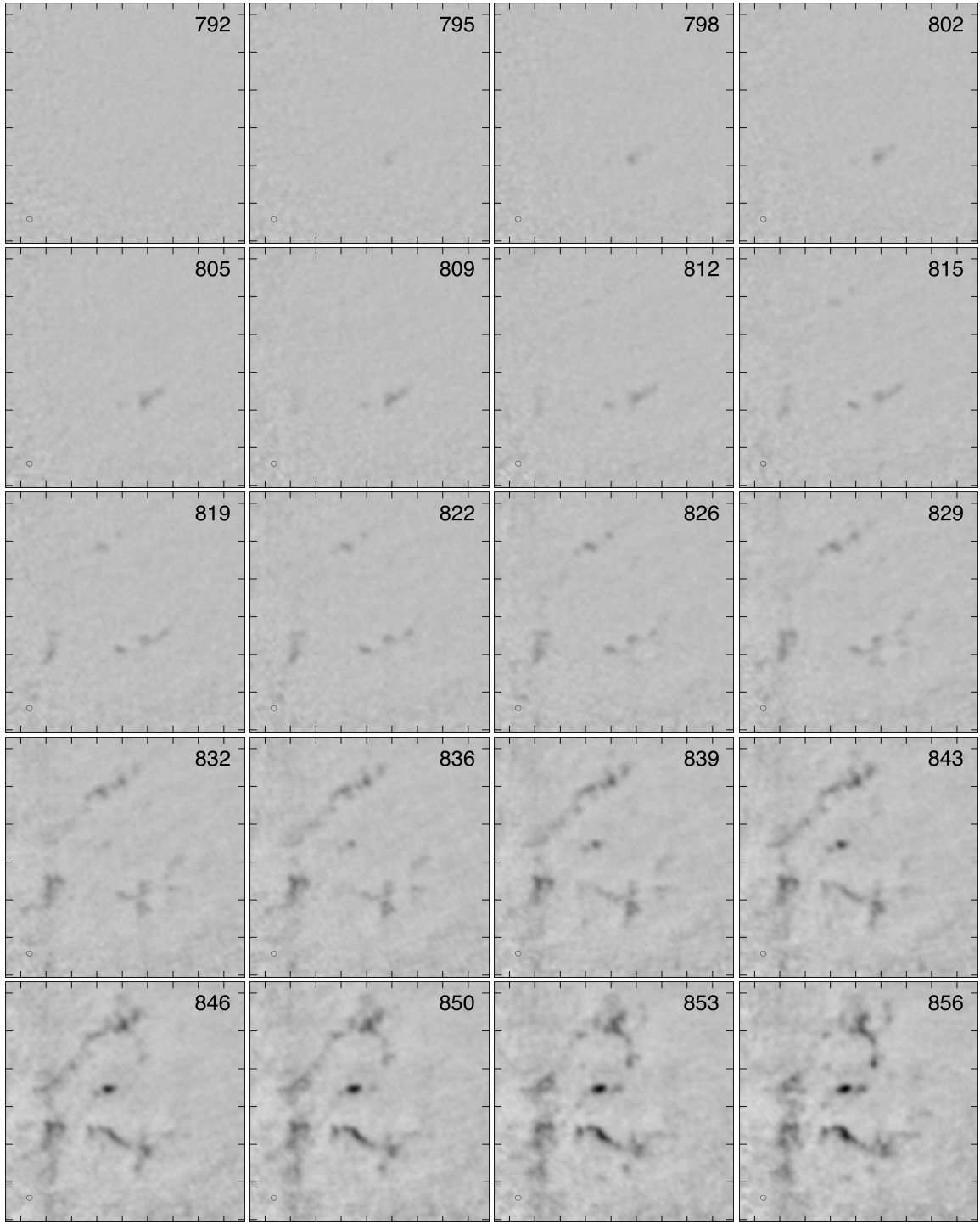} 
\caption{Velocity channel maps of the CO(3-2) emission in He 2-10, in the barycentric system.  Each figure is labelled with the velocity in \kms~.  The maps have been  Hanning-smoothed in velocity giving a channel width of 2.6\kms .  $V_{sys}$ is $872\pm1$\kms. }
\end{center}
\end{figure} 
\newpage
\includegraphics*[width=6in]{ChannelsBW_p2.pdf}
\clearpage
\includegraphics*[width=6in]{ChannelsBW_p3.pdf}

\end{document}